\documentclass[12pt]{article}

\usepackage{amsmath, amsthm, amsfonts, amssymb, bm, bbm}
\usepackage{algorithm, algpseudocode, graphicx, subfigure, epstopdf}
\usepackage{hyperref, xcolor, tabularx, multirow, booktabs, enumerate, natbib, xr, comment}

\def\spacingset#1{\renewcommand{\baselinestretch}%
{#1}\small\normalsize} \spacingset{1}

\newtheorem{Theorem}{Theorem}
\newtheorem{Assumption}{Assumption}

\newtheorem{Proposition}{Proposition}

\newcommand{\E}{\mathbb{E}}

\newcommand{\Rbb}{\mathbb{R}}

\newcommand{\base}{\mathrm{base}}
\newcommand{\tot}{\mathrm{total}}
\newcommand{\inv}{\mathrm{inv}}
\newcommand{\res}{\mathrm{res}}
\newcommand{\aug}{\mathrm{aug}}
\newcommand{\wts}{\mathrm{wts}}
\newcommand{\ISL}{\mathrm{ISL}}
\newcommand{\MTL}{\mathrm{MTL}}

\newcommand{\blind}{1}

\begin{document}

\if1\blind
{
\title{\Large{\textbf{Statistical Learning for Latent Embedding Alignment \\
with Application to Brain Encoding and Decoding}}} 
\author{
\bigskip
\large{Shuoxun Xu$^*$, Zhanhao Yan\thanks{Co-first authors}, and Lexin Li\thanks{Corresponding author}} \\
\normalsize{\textit{University of California at Berkeley}} \\
}
\date{}
\maketitle
} \fi

\if0\blind
{
\title{\Large{\textbf{Statistical Learning for Latent Embedding Alignment \\
with Application to Brain Encoding and Decoding}}}
\author{
\bigskip
\vspace{0.4in}
}
\date{}
\maketitle
} \fi

\spacingset{1.2} 
\begin{abstract}

Brain encoding and decoding aims to understand the relationship between external stimuli and brain activities, and is a fundamental problem in neuroscience. In this article, we study latent embedding alignment for brain encoding and decoding, with a focus on improving sample efficiency under limited fMRI-stimulus paired data and substantial subject heterogeneity. We propose a lightweight alignment framework equipped with two statistical learning components: inverse semi-supervised learning that leverages abundant unpaired stimulus embeddings through inverse mapping and residual debiasing, and meta transfer learning that borrows strength from pretrained models across subjects via sparse aggregation and residual correction. Both methods operate exclusively at the alignment stage while keeping encoders and decoders frozen, allowing for efficient computation, modular deployment, and rigorous theoretical analysis. We establish finite-sample generalization bounds and safety guarantees, and demonstrate competitive empirical performance on the large-scale fMRI-image reconstruction benchmark data.
\end{abstract}
\bigskip


\newpage
\spacingset{1.8} 

\section{Introduction}

Brain encoding and decoding aims to understand the relationship between external stimuli and brain activities: encoding models how the brain transforms external information into neural signals, while decoding infers and reconstructs stimuli or cognitive states from recorded neural activities. It is a fundamental problem in cognitive and computational neuroscience, as it reveals how the brain represents, processes, and interprets information, providing crucial insights into the neural mechanisms underlying perception, cognition, and behavior. It also plays a central role in the development of brain-computer-interface technologies, by identifying informative neural representations and enabling algorithms that accurately translate brain activities into intended commands for external devices. In this article, we focus on an important class of encoding and decoding problems, namely, visual reconstruction of natural image stimuli using functional magnetic resonance imaging (fMRI). In recent years, there has been a surge of research on this topic, thanks to rapid advances in neuroimaging technologies and breakthroughs in deep learning models \citep[among others]{Gaziv2022SelfSupervised, OzcelikVanRullen2023, takagi2023high, scotti2023mindeye, liu2023brainclip, chen2023seeingbeyond, GuJamisonKuceyeskiSabuncu2024, huo2024neuropictor}. See also \citet{Rakhimberdina2021Survey,Guo2025Survey} for reviews. Despite the rapid progress, however, numerous challenges remain, including the high complexity of natural images, the low signal-to-noise ratio of fMRI, the limited availability of paired natural image and fMRI samples, and substantial subject-to-subject variability. 

\begin{figure}[t!]
\centering
\includegraphics[width=0.75\textwidth,height=1.65in]{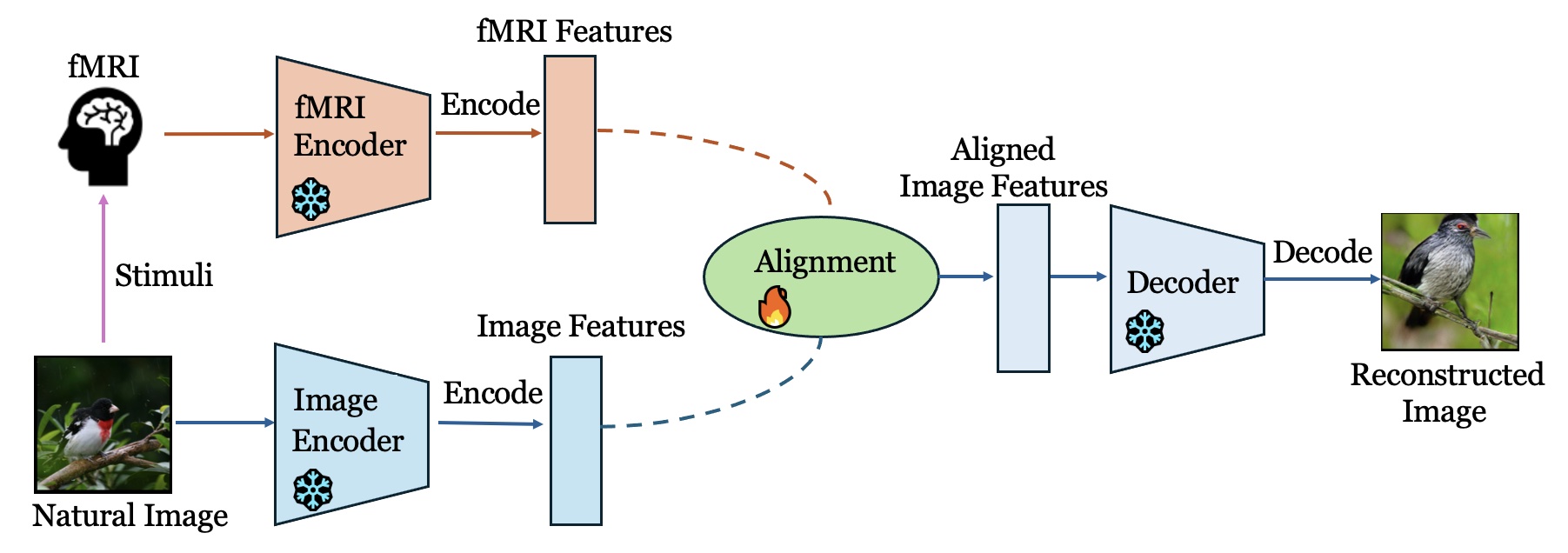}
\caption{Schematic plot of brain encoding and decoding.}
\label{fig:schematic}
\end{figure}

Visual reconstruction typically consists of three main steps: encoding, alignment, and decoding; see Figure \ref{fig:schematic} for an illustration. First, in the encoding phase, an encoder transforms raw visual stimuli such as natural images into a latent stimulus representation that captures essential structural and semantic information. In parallel, another encoder maps the fMRI signals into a latent neural representation that summarizes brain activity patterns. Encoder models built on contrastive language-image pretraining \citep[CLIP,][]{Radford2021CLIP}, vision transformer \citep[ViT,][]{Dosovitskiy2021ViT} and residual network \citep[ResNet,][]{He2016ResNet} are widely used to create such embeddings. Then, in the alignment phase, the focus is to learn a mapping from the neural latent representation to the stimulus latent representation. By establishing this mapping, one can predict the stimulus latent representation from a new fMRI recording. Finally, in the decoding phase, the predicted stimulus latent representation is translated back into the visual domain to reconstruct the perceived images. Decoder models built on generative adversarial networks \citep[GANs,][]{Goodfellow2014GAN}, variational autoencoders \citep[VAEs,][]{Kingma2014VAE}, diffusion models \citep{Ho2020DDPM} and their variants are commonly employed for this purpose. Together, encoding, alignment, and decoding form an integrated pipeline for transforming brain-derived signals into coherent and meaningful visual reconstructions.
 
In this article, we study the visual reconstruction problem by concentrating on the \emph{alignment} step, which, from a statistical learning perspective, can be formulated as a supervised regression problem. The goal is to learn a mapping from the latent representation of the fMRI recording, usually in the form of a vector, to the latent representation of the natural image stimulus, again in the form of a vector. Existing solutions include classical regression models, canonical correlation analysis, optimal transport, and more recently, contrastive learning \citep{Han2022Alignment, chen2023seeingbeyond}. Notably, \citet{qian2024lea} proposed to use a linear mapping for alignment, and showed that it achieves a reasonable reconstruction accuracy. Built on this observation, we consider a relatively simple multi-layer perceptron (MLP)-based regression model for alignment, and couple it with two statistical learning components, a new inverse semi-supervised learning approach to address the challenge of limited sample size of natural image and fMRI pairs, and a new transfer learning approach to address the challenge of substantial subject variability. For both cases, we develop statistical approaches tailored to the unique characteristics of the visual reconstruction problem. We also establish rigorous theoretical properties, and demonstrate the competitive empirical performance of our new reconstruction methods. 

Our proposal enjoys several advantages and makes useful contributions to both brain encoding and decoding as well as statistical learning in general. First of all, visual reconstruction, or more broadly, brain encoding and decoding and brain-computer-interface, represents an important application of artificial intelligence (AI). With the recent rapid advancement of AI technologies and increasingly powerful AI tools, a crucial question arises: how can classical statistical learning contribute to and facilitate this ongoing progress? The proposed work can be seen as an attempt in this direction. Specifically, we demonstrate how the integration of statistical thinking, the application of statistical principles, and the development of suitably modified statistical models can facilitate AI methodologies. Through such efforts, we aim to illustrate that statistical approaches can play a useful role in advancing AI, including in complex domains such as brain-computer-interface. Second, our proposal also advances the broader application of brain encoding and decoding. The proposed solutions are not limited to specific encoder-decoder architectures, nor particular stimulus-neural activity pairs such as natural images and fMRI, but are applicable to a wide range of encoding and decoding settings, including image, text, audio, video stimuli, and fMRI, electroencephalogram modalities. More importantly, built on relatively simple statistical principles, it offers clear computational advantages: the proposed inverse semi-supervised approach uses only one-tenth the number of free parameters of advanced alternatives, and the transfer learning approach requires roughly half as many training samples as the baseline. Although it may achieve lower reconstruction accuracy than the most sophisticated deep learning models, its empirical performance remains competitive. As such, our lightweight design offers a practically useful tradeoff, enabling more efficient deployment on platforms such as wearable or portable devices. Finally, our semi-supervised and transfer learning approaches are of independent values and contribute to general statistical methodology as well. Although inspired by classical semi-supervised and transfer learning ideas, they are \emph{not} a straightforward adaptation of existing techniques. In particular, our inverse semi-supervised learning approach reverses the roles of predictors and responses, and combines pseudo-predictor construction with residual debiasing, creating a new paradigm than traditional semi-supervised learning. Our transfer learning approach avoids pooling multi-subject data via joint fitting, but instead integrates pretrained models through sparse weighting followed by residual correction, offering a modular and privacy-preserving alternative to existing parameter-transfer strategies. Moreover, both components are supported by rigorous theoretical guarantees, including explicit upper bounds on the generalization risk, the safety properties ensuring performance no worse than the baseline, as well as the quantified gains in terms of the finite-sample generalization bound.

We also remark that, in our study, we deliberately focus on training the alignment step \emph{only}, while keeping both the image and fMRI encoders, as well as the decoder, \emph{frozen}. This design is driven by several considerations. One is computational cost, as fine-tuning large encoder or decoder models would require substantial computational resources and large amount of paired samples. In contrast, a lightweight alignment module enables fast training, low computational cost, and competitive performance under limited data. Moreover, although the encoders and decoder are kept frozen, strengthening the alignment mapping itself can still considerably improve reconstruction accuracy, since a main source of reconstruction error comes from the inaccuracy in mapping the latent features of fMRI and stimulus. Finally, the alignment step is precisely the component most amenable to statistical analysis. It has a clear supervised regression structure that allows rigorous characterization of estimation error, bias correction, and sample efficiency gains. In contrast, the encoder and decoder involve highly complicated architectures for which meaningful statistical guarantees are far more difficult to obtain. Therefore, focusing on alignment achieves a balance between practical feasibility and theoretical tractability, enabling both effective reconstruction performance and principled statistical understanding. 

The rest of the article is organized as follows. Section \ref{sec:setup} formulates the alignment problem. Section \ref{sec:semi} presents inverse semi-supervised learning for alignment. Section \ref{sec:transfer} presents transfer learning for alignment. Section \ref{sec:numericals} conducts numerical studies. Section \ref{sec:discussions} concludes with a discussion. The Supplementary Appendix collects all technical proofs.

\section{Latent Embedding Alignment Setup}
\label{sec:setup}

In a typical brain encoding-decoding experiment using natural image stimuli, participants view a sequence of images while their brain activity is recorded using fMRI, a noninvasive neuroimaging technique that measures brain activity indirectly through the blood-oxygen-level-dependent (BOLD) signal. Each fMRI image contains tens of thousands of voxels, and each voxel summarizes the aggregated activities of thousands of neurons in a small cortical region, sampled at regular time intervals, typically every 0.5 to 2 seconds. During the experiment, natural images, ranging from objects to complex scenes, are presented in a controlled manner, whereas fMRI signals are collected. To connect the external visual input with the internal neural representation, both the visual stimuli and the brain neural responses are transformed into latent feature representations through separate encoders. A natural image encoder, such as a convolutional neural network, e.g., ResNet, or a transformer-based vision model, e.g., ViT or CLIP, extracts hierarchical visual features capturing structural, semantic, and contextual information from each image. These features form a vector in a high-dimensional latent space that summarizes the key visual content. Similarly, an fMRI encoder maps the raw voxel-wise BOLD activations into another latent vector that captures the spatial pattern of neural activities across different regions. 

Let $X \in \Rbb^d$ and $Y \in \Rbb^m$ denote the latent feature vectors from the encoders of neural activities and stimuli, respectively. The typical latent feature dimensions $d, m$ are 512, 1024, 2048, and in our experiment, we choose $d = 2048$, and $m = 1024$. To formalize the alignment problem, we view it as a supervised learning task in which the goal is to construct a predictive mapping from the latent fMRI feature vector $X$ to the latent visual feature vector $Y$. Specifically, define the mapping, 
\begin{equation} \label{eqn:model}
f^* : \Rbb^d \to \Rbb^m, \; \; f^*(X) = \E(Y | X), 
\end{equation}
which is the main target of interest in our alignment problem. Define the noise, 
\begin{equation} \label{eqn:noise}
e = Y - f^*(X), \; \textrm{ such that } \; 
\mathbb{P}\left(\left|\langle e, u \rangle\right| > t\right) \leq 2\exp\left(-\frac{t^2}{2\sigma^2}\right), 
\end{equation}
for any unit vector $u \in \mathbb{R}^m$ and any $t > 0$. This noise term captures various sources of discrepancy, including encoder mismatch, measurement variability, and latent structure missed by the mapping $f^*$. It is assumed to have zero mean, and is sub-Gaussian with parameter $\sigma^2$ following \eqref{eqn:noise}, which is a quite mild condition. 

We learn $f^*$ via a multi-layer perceptron \citep[MLP,][]{rumelhart1986learning}, 
\vspace{-0.01in}
\begin{equation*}
f_{\Theta}(X) = \theta_L f_L\big( \theta_{L-1} \cdots f_1(\theta_0 X) \cdots \big),
\end{equation*}
indexed by the parameter space, 
\begin{equation*}
\mathcal{M} = \left\{ \Theta = (\theta_L, \ldots, \theta_0) : \theta_l \in \mathbb{R}^{p_{l+1} \times p_l}, d = p_0, m = p_{L+1} \right\},
\end{equation*}
where $p_0 = d$ is the input dimension, $p_{L+1} = m$ is the output dimension, $p_{\max} = \max_{l \in \{0,\ldots,L-1\}}$ $p_{l+1}$ is the maximum network width, $p_{\tot} = \sum_{l=0}^L p_{l+1} p_l$ is the total number of parameters, and $L$ is the network depth. In our implementations, $L = 4$ or $5$. The activation functions $f_l : \mathbb{R} \to \mathbb{R}$ are applied element-wise and are assumed to be 1-Lipschitz with $f_l(0) = 0$. This is satisfied by many common activation functions such as ReLU, leaky ReLU, hyperbolic tangent, and sigmoid, and by incorporating bias terms into the weight matrices.

We further introduce regularizations. Consider $q \in [1,2]$, and for any matrix $\theta_l \in \mathbb{R}^{p_{l+1} \times p_l}$, we define the entry-wise $\ell_q$ norm as $\|\theta_l\|_q = (\sum_{i=1}^{p_{l+1}} \sum_{j=1}^{p_l} |(\theta_l)_{ij}|^q )^{1/q}$. The choice of $q \in [1,2]$ encompasses the $\ell_1$ norm with $q=1$ that promotes element-wise sparsity, and the $\ell_2$ norm with $q=2$ that favors a smooth representation and stabilizes estimation. Besides, it helps maintain the theoretical tractability and Lipschitz control. This is because, for any $q \in [1,2]$, we have $\|\theta_l\|_{\text{op}} \leq \|\theta_l\|_2 \leq \|\theta_l\|_q \leq \|\theta_l\|_1$, where $\|\theta_l\|_{\text{op}}$ denotes the operator norm. As such, constraining $\|\theta_l\|_q \leq 1$ automatically bounds the operator norm by 1, which is critical for controlling the Lipschitz constant of the neural network: $\text{Lip}(f_{\theta}) \leq \prod_{l=0}^L \|\theta_l\|_{\text{op}}$. Under regularization, we focus on the constrained parameter space, 
\begin{equation*}
\mathcal{M}_q = \left\{\Theta \in \mathcal{M} : \max_{l \in \{0,\ldots,L\}} \|\theta_l\|_q \leq 1\right\}.
\end{equation*}
The theoretical results derived later hold uniformly for all $q \in [1,2]$, and can be easily extended to any parameter space with a bounded entry-wise $\ell_q$ norm. 

Before turning to semi-supervised and transfer learning, we first consider the baseline alignment method. For a given participant, suppose we observe $n$ i.i.d.\ copies $\{ (X_i,Y_i), i=1,\ldots,n \}$ of $(X,Y)$, where $n$ denotes the number of natural images shown to this participant while the brain activity is recorded by fMRI. We consider
\vspace{-0.01in}
\begin{equation}\label{eqn:baseline}
\widehat{\Theta}_{\base} = \arg\min_{\Theta \in \mathcal{M}_q} \frac{1}{n}\sum_{i=1}^n \left\|Y_i  - f_{\Theta}({X}_i)\right\|_2^2 + \lambda_\base \|\theta_L\|_q^q,
\end{equation}
where the regularization is placed on the last layer parameter $\theta_L \in \mathbb{R}^{p_{L+1}\times p_L}$. This formulation simplifies the subsequent theoretical analysis, while it can be extended straightforwardly to regularizations involving all layers of parameters, such as $\lambda \sum_{l=0}^L \|\theta_l\|_q^q$ or $\lambda \prod_{l=0}^L \|\theta_l\|_q$, following similar analytical techniques of \citet{lederer2024statistical}. We have the following generalization bound. Let $\Theta^* = (\theta_L^{*}, \ldots, \theta_0^{*}) = \arg\min_{\Theta\in\mathcal{M}_q}\mathbb{E}\|f^*(X) - f_{\Theta}(X)\|_2^2$ denote the population risk minimizer. For two sequences $a$ and $b$, we write $a \lesssim b$ if there exists a positive constant $C$ such that $a \leq C b$, and write $a \asymp b$ if $a \lesssim b$ and $b \lesssim a$.

\begin{Proposition}\label{Naive-Gen-Bound}
Suppose $\lambda_\base \asymp v_{\infty} \sqrt{{L \{ \log(2mnp_{\tot}) \}^3}/{n}}$, where $v_{\infty} = \sqrt{n^{-1}\sum_{i=1}^n \|X_i\|_\infty^2}$. Then, with probability approaching 1 as $n \to \infty$, we have 
\begin{align*} 
\mathbb{E}\|f_{\widehat{\Theta}_{\base}}(X) - f^*(X)\|_2^2 \lesssim &  \underbrace{\inf_{\Theta\in\mathcal{M}_q} \mathbb{E}\|f_{\Theta}(X) - f^*(X)\|_2^2}_{\mathcal{E}_{\base, 1}} \notag\\
       & + \underbrace{v_\infty \sqrt{\frac{L\{ \log(2mn p_{\tot}) \}^3}{n}} \|\theta_L^{*}\|_q^q}_{\mathcal{E}_{\base, 2}} +  \underbrace{v_\infty^2 \frac{L\{ \log(2mn p_{\tot}) \}^3}{n}}_{\mathcal{E}_{\base, 3}}.
\end{align*}
\end{Proposition}

Proposition \ref{Naive-Gen-Bound} decomposes the generalization error of the baseline method into three components: the approximation error $\mathcal{E}_{\base, 1}$ that measures how well the model class $\mathcal{M}_q$ approximates the true function $f^*$, the statistical error $\mathcal{E}_{\base, 2}$ that depends on the input scale $v_\infty$, the network complexity $\sqrt{L\{\log(2mn p_{\tot})\}^3}$, and the model complexity measure $\|\theta_L^{*}\|_q^q$, and the higher-order term $\mathcal{E}_{\base, 3}$.

\section{Semi-supervised Learning for Alignment}
\label{sec:semi}

\subsection{Methodology}
\label{sec:method-semi}

We first consider semi-supervised learning. In a typical experiment, the number of available fMRI-image pairs $n$ is limited, typically ranging around a few thousands. In our dataset, $n$ is around $8,000$. The main reason is that collecting fMRI data is costly, time-consuming, and constrained by practical considerations, for instance, each participant can only undergo a relatively small number of scanning sessions. In contrast, natural images are abundant and easily obtainable in virtually unlimited quantities. This severe imbalance between scarce neural data and plentiful visual data motivates us to develop a method that can effectively leverage the vast corpus of natural images to enhance encoding-decoding performance.

This problem is related to semi-supervised learning, which aims to improve model performance by leveraging both labeled and unlabeled data \citep{deng2024optimal, cai2025semi}. In classical semi-supervised learning settings, a large number of feature observations $X$ are available, but only a small subset have corresponding labels $Y$, and the goal is to use the structure of the abundant unlabeled $X$ samples to guide the learning when labeled pairs $(X,Y)$ are scarce. Our setting is similar in spirit, but significantly differs, in that the roles between features and responses are \emph{reversed}. Specifically, the number of samples of response $Y$, i.e., the latent features of natural images, is vast, whereas the number of samples of predictor $X$, i.e., the latent features of fMRI measurements, is limited. 

To address this challenge, we develop an \emph{inverse semi-supervised learning} (ISL) approach. In addition to the paired observations $(X_i, Y_i)$ for $i=1,\ldots,n$, suppose we have additional unpaired response features $Y_i$ for $i=n+1, \ldots, n+N$, generated through the image encoder with $N$ additional natural images that are not shown to the participant with no corresponding fMRI recording. Our goal is to learn the mapping $f^*$ in \eqref{eqn:model} by leveraging both the paired data $(X_i, Y_i)$, which are limited in quantity,  and the unpaired data $Y_i$, which are abundant. Our method consists of three main steps. 

In the first step, we regress $X_i$ on $Y_i$ for $i=1,\ldots,n$ to learn an inverse mapping $\widehat{g} : \Rbb^m \to \Rbb^d$, then construct the pseudo-predictors $\widehat{g}(Y_i)$ for $i=n+1, \ldots, n+N$ based on the learnt inverse mapping $\widehat{g}$. That is, we obtain $\widehat{g} = f_{\widehat{\Theta}_1}$ by fitting an MLP, where 
\vspace{-0.01in}
\begin{equation} \label{eqn:isl-step1}
\widehat{\Theta}_\inv = \arg\min_{\Theta \in \mathcal{M}_q} \frac{1}{n}\sum_{i=1}^n \left\|X_i  - f_{\Theta}(Y_i) \right\|_2^2 + \lambda_{\inv} \|\theta_L\|_q^q
\end{equation}

In the second step, we regress $Y_i$ on $X_i$ for $i=1,\ldots,n$ and regression $Y_i$ on $\widehat{g}(Y_i)$ for $i=n+1, \ldots, n+N$ to learn an augmented model with both observed and pseudo-features, by fitting another MLP, 
\vspace{-0.01in}
\begin{equation} \label{eqn:isl-step2}
\widehat{\Theta}_\aug = \arg\min_{\Theta \in \mathcal{M}_q} \frac{1}{n+N}\left\{ \sum_{i=1}^{n}\|Y_i - f_\Theta(X_i)\|_2^2 + \sum_{i=n+1}^{n+N}\|Y_i - f_\Theta(\widehat{g}(Y_i))\|_2^2 \right\} + \lambda_{\aug} \|\theta_L\|_q^q.
\end{equation}

In the third step, we compute the residual $\{Y_i - f_{\widehat{\Theta}_\aug}(X_i)\}$ from the learnt augmented model $f_{\widehat{\Theta}_\aug}$, and regress the residual on $X_i$ for $i=1,\ldots,n$ to learn a residual correction,
\begin{equation} \label{eqn:isl-step3}
\widehat{\Theta}_\res = \arg\min_{\Theta \in \mathcal{M}_q} \frac{1}{n}\sum_{i=1}^n\|\{ Y_i - f_{\widehat{\Theta}_\aug}(X_i) \} - f_\Theta(X_i)\|_2^2 + \lambda_{\res} \|\theta_L\|_q^q.
\end{equation}

Finally, we obtain our estimated mapping, 
\vspace{-0.05in}
\begin{equation} \label{eqn:isl-final}
\widehat{f}_{\ISL}(X) = f_{\widehat{\Theta}_\aug}(X) + f_{\widehat{\Theta}_\res}(X).
\vspace{-0.05in}
\end{equation} 
Algorithm \ref{alg:isl} summarizes the above estimation procedure. 

\begin{algorithm}[t!] 
\caption{Inverse semi-supervised learning procedure.}
\begin{algorithmic}[1]
\State \textbf{Input}: Observed paired data $\{ (X_i, Y_i), i=1,\ldots,n \}$, and unpaired augmentation data $\{Y_i, i=n+1,\ldots,n+N\}$.
\State \textbf{Step 1}: Learn the inverse mapping $\widehat{g}$ via \eqref{eqn:isl-step1}, by regressing $X_i$ on $Y_i$, $i=1,\ldots,n$. 
\State \textbf{Step 2}: Learn the augmented model with both observed and pseudo-features via \eqref{eqn:isl-step2}, by regressing $Y_i$ on $X_i$, $i=1,\ldots,n$, and regressing $Y_i$ on $\widehat{g}(Y_i)$, $i=n+1, \ldots, n+N$. 
\State \textbf{Step 3}: Learn the residual via \eqref{eqn:isl-step3}, by regressing $\{Y_i - f_{\widehat{\Theta}_\aug}(X_i)\}$ on $X_i$, $i=1,\ldots,n$. 
\State \textbf{Output}: $\widehat{f}_{\ISL}(X) = f_{\widehat{\Theta}_\aug}(X) + f_{\widehat{\Theta}_\res}(X)$ as in \eqref{eqn:isl-final}.
\end{algorithmic}
\label{alg:isl}
\end{algorithm}

We next make some remarks regarding our proposed ISL method. First, ISL is specifically designed to leverage the abundance of natural images while preserving robustness when the inverse mapping $\widehat{g}$ is imperfect. When the unpaired response features $\{Y_i, i=n+1, \ldots, n+N\}$ are informative and $\widehat{g}$ provides a reasonable approximation of the inverse mapping, Step 2 effectively increases the sample size from $n$ paired observations to $n+N$ augmented observations. Even though the pseudo-predictors $\widehat{g}(Y_i)$ may be noisy or biased, their inclusion still provides additional information about the forward mapping. 

Second, because the inverse mapping inevitably introduces bias, ISL incorporates a residual fitting step to explicitly correct this bias. The key idea is that the discrepancy between the estimator $f_{\widehat{\Theta}_\aug}(X_i)$ in Step 2 and the true mapping $f^*$ reflects systematic error induced by imperfect pseudo-features. This bias manifests as the residual
$\{Y_i - f_{\widehat{\Theta}_\aug}(X_i)\}$ that is learnable from the paired data. By regressing this residual on $X_i$, Step 3 captures a portion of $f^*$ that Step 2 fails to learn due to noisy or biased pseudo-features. The final estimator $f_{\widehat{\Theta}_\aug}(X) + f_{\widehat{\Theta}_\res}(X)$ then cancels the bias introduced by pseudo-predictors. This residual-based bias correction strategy has been used in semi-supervised learning \citep{deng2024optimal, cai2025semi} and in high-dimensional penalized regressions \citep{raskutti2009minimax}. In our setting, residual fitting ensures that even if $\widehat{g}$ is inaccurate, the final ISL estimator is never worse than the baseline estimator trained without any additional data. 

Third, ISL differs fundamentally from classical semi-supervised learning. Standard methods assume abundant unlabeled predictors and use their distributional structure to regularize the supervised task. In contrast, our setting reverses the roles: we have abundant responses and need to construct pseudo-predictors through an inverse mapping. This inversion changes the nature of the bias introduced by unlabeled data and requires bias correction to guarantee valid statistical behavior. Moreover, our approach is built around nonlinear MLP models, going beyond the linear or kernel-based models commonly considered in classical theory of semi-supervised learning \citep{deng2024optimal, cai2025semi}. 

Taken together, ISL offers a statistically principled and robust way to improve alignment estimation using abundant unpaired natural images. It enjoys performance enhancement when the inverse mapping is informative, and a built-in safeguard when the inverse mapping is noisy or uninformative. We next establish rigorous characterizations of these observations through formal theoretical analyses.

\subsection{Theoretical analysis}
\label{sec:theory-semi}

We first derive the explicit upper bound on the generalization risk. We then show that the ISL method provably improves and never performs worse than the baseline method. 

We begin with two regularity conditions on the local geometry of the loss landscape, and the quality of auxiliary information. 

\begin{Assumption}[Local quadratic growth]
\label{assump:quadratic-growth}
Suppose there exists $\mu > 0$, such that, for all $\Theta$ in a neighborhood of the population risk minimizer $\Theta^*$, $\mathbb{E}\|f_{\Theta}(X) - f^*(X)\|_2^2 - \mathbb{E}\|f_{\Theta^*}(X) - f^*(X)\|_2^2 \geq \mu \mathbb{E}\|f_{\Theta}(X) - f_{\Theta^*}(X)\|_2^2$. 
\end{Assumption}

This assumption characterizes the local geometry of the population risk near its minimizer. It states that the excess risk is lower bounded by the squared prediction difference, ensuring that the parameters close in risk are also close in prediction, with a larger $\mu$ implying a faster convergence. Unlike global strong convexity, it only requires the condition to hold in a neighborhood of $\Theta^*$, making it much weaker and more realistic in practice. It is satisfied for overparameterized neural networks under mild conditions, and is a standard condition in deep learning theory \citep{du2019gradient, allen2019convergence}. It is important for converting generalization bounds on the excess risk into convergency rates for the prediction error, and in our context enables us to derive the rate for $\mathbb{E}\|f_{\widehat{\Theta}}(X) - f_{\Theta^*}(X)\|_2^2$.

\begin{Assumption}[Inverse mapping quality]
\label{assump:inverse}
For ISL, let $g^*(Y) = \E(X | Y)$ denote the true inverse mapping. Suppose the estimated inverse mapping $\widehat{g}$ from Step 1 of ISL satisfies that $\E\|\widehat{g}(Y) - g^*(Y)\|_2^2 \leq C_{\text{inv}}$, for some constant $C_{\text{inv}} > 0$.
\end{Assumption}

This assumption states that the inverse mapping learned from paired data has bounded mean squared error. It is a mild condition, in that it only requires the inverse mapping error not to be arbitrarily inaccurate.

Next, we establish a finite-sample generalization bound that explicitly quantifies how unlabeled data enhances learning through the inverse mapping mechanism. Let $\Theta^*_{\aug} = (\theta^*_{L,\aug}, \ldots, \theta^*_{0,\aug}) = \arg\min_{\Theta\in\mathcal{M}_q} n (n+N)^{-1} \mathbb{E}\|f^*(X) - f_{\Theta}(X)\|_2^2 + N (n+N)^{-1} \mathbb{E}\|Y - f_{\Theta}(\widehat{g}(Y))\|_2^2$, and $\Theta^*_{\res} = (\theta^*_{L,\res}, \ldots, \theta^*_{0,\res}) = \arg\min_{\Theta\in\mathcal{M}_q}\mathbb{E}\|\{f^*(X) - f_{\Theta^*_{\aug}}(X)\} - f_{\Theta}(X) \|_2^2$ denote the population risk minimizer corresponding to Steps 2 and 3 of ISL, respectively. 

\begin{Theorem}[Generalization bound for ISL] \label{thm:isl}
Suppose Assumptions \ref{assump:quadratic-growth} and \ref{assump:inverse} hold. Suppose $q \in [1,2]$, $\lambda_{\inv} \asymp v_{Y,\infty}\sqrt{{L\{\log(ndp_{\tot})\}^3}/{n}}$, $\lambda_{\aug} \asymp v_\infty\sqrt{{L[\log\{(n+N)mp_{\tot}\}]^3}/{(n+N)}}$, and $\lambda_{\res} \asymp v_\infty\sqrt{{L\{\log(nmp_{\tot})\}^3}/{n}}$, where $v_{Y,\infty} = \sqrt{n^{-1} \sum_{i=1}^n \|Y_i\|_\infty^2}$. Then, with probability approaching 1 as $n \to \infty$, we have
\vspace{-0.01in}
\begin{align*}
\E\|\widehat{f}_{\ISL}(X) - f^*(X)\|_2^2 \lesssim & \underbrace{\inf_{\Theta\in\mathcal{M}_q} \mathbb{E}\|f_{\Theta}(X) - \{ f^*(X) - f_{\Theta^*_{\aug}}(X) \} \|_2^2}_{\text{$\mathcal{E}_{\ISL,1}$: Residual space approximation error}} \notag\\
&\quad + \underbrace{v_\infty \sqrt{\frac{L[ \log\{ (n+N)mp_{\tot}\} ]^3}{n+N}} \|\theta^*_{L,\aug}\|_q^q}_{\text{$\mathcal{E}_{\ISL,2}$: Augmented learning statistical error}}  \notag\\
&\quad + \underbrace{v_\infty \sqrt{\frac{L\{ \log( nmp_{\tot} ) \}^3}{n}} \|\theta^*_{L, \res}\|_q^q}_{\text{$\mathcal{E}_{\ISL,3}$: Residual fitting statistical error}}  + \underbrace{v_\infty^2 \frac{L\{ \log(nmp_{\tot}) \}^3}{n}}_{\text{$\mathcal{E}_{\ISL,4}$: Higher-order term}}.
\end{align*}
\end{Theorem}

Theorem \ref{thm:isl} shows that ISL achieves a finite-sample bound where the total error decomposes into four terms: a residual space approximation error, two statistical errors corresponding to the steps of augmented learning and residual fitting, and a higher-order term. Specifically, the first term, $\mathcal{E}_{\ISL,1}$, represents the approximation error in the residual space, which measures how well the neural network class $\mathcal{M}_q$ approximates the residual $f^*(X) - f_{\Theta^*_{\aug}}(X)$. When the augmented learning step successfully captures information in $f^*$, this error term decreases. A key insight here is that the quality of the inverse mapping and the quantity of unlabeled data jointly determine the complexity of the residual space, which in turn affects the final accuracy. The second term, $\mathcal{E}_{\ISL,2}$, quantifies the statistical error in the augmented learning step using the combination of $n$ paired data and $N$ unlabeled data. It reflects a direct benefit of utilizing unlabeled data, effectively increasing the sample size from $n$ to $n + N$. The third term, $\mathcal{E}_{\ISL,3}$, quantifies the statistical error in the residual fitting step, which uses only the $n$ labeled samples. Unlike $\mathcal{E}_{ISL,2}$, it reflects the indirect benefit of using unlabeled data. That is, when the augmented learning captures sufficient information in $f^*$, learning the residual leads to a smaller error than learning $f^*$ directly. The fourth term, $\mathcal{E}_{\ISL,4}$, is a higher-order term that vanishes faster than the rest, and becomes negligible when the sample size is sufficiently large. We give the proof of Theorem \ref{thm:isl} and discuss its main challenge in Appendix Section S2.2. 

Finally, we establish the performance improvement and safeguard property of ISL.

\begin{Theorem}[Performance guarantee of ISL]
\label{cor:isl-enhancement}
Suppose the conditions of Theorem \ref{thm:isl} hold. 
\begin{enumerate}[(a)]
\item (Safety). When $\|\theta^*_{L,\res}\|_q^q \lesssim \|\theta^*_L\|_q^q$, ISL is never worse than the baseline method, in that
\begin{equation*}
\E\|\widehat{f}_{\ISL}(X) - f^*(X)\|_2^2 \lesssim \E\|f_{\widehat{\Theta}_{\base}}(X) - f^*(X)\|_2^2 + v_\infty \sqrt{\frac{L[ \log\{(n+N)mp_{\tot}\} ]^3}{n+N}} \|\theta^*_{L,\aug}\|_q^q.
\end{equation*}

\item (Enhancement). When (i) there is sufficient unlabeled data, in that  $\exp(n) \succ N \gtrsim n {\left( \|\theta^*_{L,\aug}\|_q^q \right)^2} / {(\|\theta_L^{*}\|_q^q)^2}$, and (ii) the inverse mapping is reasonably accurate, in that  $\E\|\widehat{g}(Y) - g^*(Y)\|_2^2 \lesssim n \{(\|\theta_L^{*}\|_q^q)^2 - (\|\theta^*_{L,\res}\|_q^q)^2\} / \{N (\|\theta^*_{L,\aug}\|_q^q)^2\}$. then ISL achieves a strict improvement over the baseline method, in that, for some constant $c > 0$, 
\vspace{-0.01in}
\begin{align*}
\E\|\widehat{f}_{\ISL}(X) - f^*(X)\|_2^2 \lesssim & \; \E\|f_{\widehat{\Theta}_{\base}}(X) - f^*(X)\|_2^2 \\ 
& \quad - c \,  v_\infty \sqrt{\frac{L\{ \log(nmp_{\tot}) \}^3}{n}} \left( \|\theta_L^{*}\|_q^q - \|\theta^*_{L,\res}\|_q^q \right). 
\end{align*}
\end{enumerate}
\end{Theorem}

Theorem \ref{cor:isl-enhancement} quantifies the performance enhancement that ISL achieves over the baseline method, and characterizes the precise conditions under which this  enhancement occurs. 

The safety guarantee shows that, even when the augmented learning step fails to improve, ISL incurs only a controlled excess risk that diminishes as $N$ increases. This property distinguishes ISL from many classical semi-supervised learning approaches that can perform worse than the baseline method when some distributional assumptions are violated \citep{singh2008unlabeled, nadler2009statistical}. In ISL, the worst-case degradation is bounded by the statistical error in the augmented learning step, which vanishes at rate $1/\sqrt{n+N}$ and becomes negligible when $N \gg n$. The condition that $\|\theta^*_{L,\res}\|_q^q \lesssim \|\theta^*_L\|_q^q$ to ensure such a safety essentially requires that  
the complexity of the residual learning does not exceed that of directly learning $f^*$, which is a fairly mild and reasonable requirement. 

The enhancement guarantee shows that, the improvement is proportional to the complexity decrease from $\|\theta_L^{*}\|_q^q$ to $\|\theta^*_{L,\res}\|_q^q$, scaled by the statistical rate $\sqrt{L\{ \log(nmp_{\tot}) \}^3/n}$. This decrease reflects the simplification of the learning problem achieved by the augmented learning step, i.e., when the augmented learning successfully captures information in $f^*$, the residual learning is to become much simpler than directly learning $f^*$ from scratch. Moreover, this is done through a mechanism different from traditional semi-supervised learning \citep{deng2024optimal, cai2025semi}. Rather than assuming that unlabeled predictors lie near the decision boundary or follow a specific distribution, ISL leverages unlabeled responses to increase the effective sample size, then simplifies the supervised learning problem in the residual space. Such an  enhancement materializes when the quantity of unlabeled data $N$ is large enough, and the inverse mapping estimation $\widehat{g}$ is reasonably accurate. 

\section{Transfer Learning for Alignment}
\label{sec:transfer}

\subsection{Methodology}
\label{sec:method-transfer}

We next turn to transfer learning. In a typical encoding-decoding experiment, model training is often done for one participant at a time, mostly due to substantial inter-subject variability in cortical organization and neural response patterns. Nevertheless, there also exist meaningful similarities across subjects, especially in the functional architecture of visual and associative cortices. This suggests that information from models trained on previous participants can be transferred to benefit the training of a new subject's model. By borrowing knowledge across subjects, it becomes possible to improve statistical efficiency and generalization, and can enable the new subject's model to achieve comparable decoding accuracy while requiring substantially fewer training samples.

Transfer learning aims to leverage information from source domains to improve learning in a target domain. Existing approaches can be broadly divided into two categories, parameter transfer and representation transfer \citep{zhu2025TLreview}. Parameter transfer, or model-based methods, such as those developed by \cite{cai2021transfer, li2022transfer, cai2024transfer}, combine and regularize model parameters across domains, usually through sparse aggregation, debiasing, or adaptive weighting, to borrow information from multiple pretrained models. Representation transfer, or data-distribution-based methods, such as those by \cite{xu2025representation, yuan2025optimal}, seek a shared low-dimensional latent structure, usually by aligning or adapting feature representations, to capture commonalities and mitigate distributional shifts between source and target domains.

We adopt the parameter transfer idea and propose a \emph{meta transfer learning} (MTL) approach, with modifications to existing solutions. Specifically, for the target subject, i.e., the new subject for whom we aim to train a model, suppose we observe $n$ i.i.d.\ copies $\{ (X_i,Y_i), i=1,\ldots,n \}$ of $(X,Y)$. Meanwhile, suppose there are $K$ source subjects, i.e., the previous subjects with their alignment models already trained, and let $\widetilde{\Theta}_{1}, \ldots, \widetilde{\Theta}_{K}$ denote the corresponding trained model parameters. Our goal is to learn the true mapping $f^*$ in \eqref{eqn:model} given the target subject's data, while leveraging the knowledge from the source subjects. Our new method consists of two main steps. 

In the first step, we feed $X_i$ of the target subject into the learnt models to construct a set of pseudo-predictors $f_{\widetilde{\Theta}_k}(X_i)$, then regress $Y_i$ on $f_{\widetilde{\Theta}_k}(X_i)$, with a Lasso type sparsity regularization, to learn a set of weights $\gamma = (\gamma_1,\ldots,\gamma_K)^T \in \Rbb^{K}$; i.e.,  
\vspace{-0.01in}
\begin{equation} \label{eqn:mtl-step1}
\widehat{\gamma} = \arg\min_{\gamma\in\mathbb{R}^K} \frac{1}{n}\sum_{i=1}^n \Big\| Y_i - \sum_{k=1}^K \gamma_k f_{\widetilde{\Theta}_k}(X_i) \Big\|_2^2 + \lambda'_{\wts} \|\gamma\|_1.
\end{equation}

In the second step, we compute the residual $\{Y_i - \sum_{k=1}^K \widehat{\gamma}_k f_{\widetilde{\Theta}_k}(X_i)\}$, and regress the residual on $X_i$ to learn a residual model; i.e., \begin{equation} \label{eqn:mtl-step2}
\widetilde{\Theta} = \arg\min_{\Theta \in \mathcal{M}_q} \frac{1}{n}\sum_{i=1}^n \Big\| \Big\{Y_i - \sum_{k=1}^K \widehat{\gamma}_k f_{\widetilde{\Theta}_k}(X_i)\Big\} - f_\Theta({X}_i) \Big\|_2^2 + \lambda'_{\res}\|\theta_L\|^q_q.
\end{equation}

Finally, we obtain our estimated mapping, 
\vspace{-0.01in}
\begin{equation} \label{eqn:mtl-final}
\widehat{f}_{\MTL}(X) = f_{\widetilde{\Theta}}(X) + \sum_{k=1}^K \widehat{\gamma}_k f_{\widetilde{\Theta}_k}(X).
\end{equation}
Algorithm \ref{alg:mtl} summarizes the estimation procedure. 

\begin{algorithm}[t!]
\caption{Meta transfer learning procedure.}
\begin{algorithmic}[1]
\State \textbf{Input}: Target data $\{(X_i,Y_i), i=1,\ldots,n\}$, and the learnt source models $\widetilde{\Theta}_1,\ldots,\widetilde{\Theta}_K$.
\State \textbf{Step 1}: Learn the sparse source information weights via \eqref{eqn:mtl-step1}, by regressing $Y_i$ on $f_{\widetilde{\Theta}_k}(X_i)$, $k=1,\ldots,K$, $i=1,\ldots,n$.
\State \textbf{Step 2}: Learn the residual on target data via \eqref{eqn:mtl-step2}, by regressing the residual $\{Y_i - \sum_{k=1}^K \widehat{\gamma}_k f_{\widetilde{\Theta}_k}(X_i)\}$ on $X_i$, $i=1,\ldots,n$. 
\State \textbf{Output}: $\widehat{f}_{\MTL}({X}) = f_{\widetilde{\Theta}}(X) + \sum_{k=1}^K \widehat{\gamma}_k f_{\widetilde{\Theta}_k}(X)$.
\end{algorithmic}
\label{alg:mtl}
\end{algorithm}

We again make some remarks regarding our proposed MTL method. First, similar to ISL, MTL is also designed to exploit auxiliary information, in this case, pretrained models from other subjects, while ensuring robustness when such information is noisy or only partially relevant. When the source models carry useful information shared by the target subject, Step 1 functions as a sparse aggregation, which adaptively selects and weighs the most informative source subjects. Because this step is a low-dimensional Lasso regression, its statistical complexity is dramatically smaller than that of fitting a new high-dimensional MLP from scratch, allowing information to be borrowed across subjects while improving the stability of the alignment estimation when the target sample size is limited.

Second, MTL also incorporates a residual fitting step to correct for systematic mismatches between the aggregated predictions from source models and the true mapping for the target subject. Step 2 fits an MLP to map $X_i$ to the residual $\{Y_i - \sum_{k=1}^K \widehat{\gamma}_k f_{\widetilde{\Theta}_k}(X_i)\}$, ensuring that the bias induced by irrelevant, inaccurate, or misspecified source models is accounted for. As a result, the final estimator $f_{\widetilde{\Theta}}(X) + \sum_{k=1}^K \widehat{\gamma}_k f_{\widetilde{\Theta}_k}(X)$ carries useful information from the source models while safeguarding against negative transfer.

Third, our strategy differs significantly from existing transfer learning approaches via parameter transfer or joint training \citep{cai2021transfer, li2022transfer, cai2024transfer}. In particular, our method operates entirely at the model-prediction level, without accessing the raw source data. This offers important practical advantages: it avoids privacy or data-sharing concerns, reduces computational burden, and allows the procedure to be applied even when pretrained source models are provided as blackboxes. Moreover, the sparse weighting mechanism  adapts to the heterogeneity of source subjects, up-weighting similar subjects and down-weighting dissimilar ones, \emph{without} requiring explicit similarity modeling.

Taken together, MTL provides a statistically principled and flexible way to leverage information across participants, while remains robust by ensuring that the performance is never worse than training on the target subject alone.

\subsection{Theoretical analysis}
\label{sec:theory-transfer}

Parallel to ISL, we first derive the explicit upper bound on the generalization risk for the proposed MTL, then establish its safety guarantee and the performance enhancement. 

In addition to Assumption \ref{assump:quadratic-growth}, we introduce two assumptions for MTL. 

\begin{Assumption}[Source model quality] \label{assump:auxiliary}
Let $\gamma^* = (\gamma_1^*,\dots,\gamma_K^*)^T \in \mathbb{R}^K = \arg\min_{\gamma} E\| f^*(X)$ $- \sum_{k=1}^K\gamma_k f_{\widetilde{\Theta}_k} \|^2$, and let $f_{\res}(X) = f^*(X) - \sum_{k=1}^K \gamma_k^* f_{\widetilde{\Theta}_k}(X)$. Suppose $\mathbb{E}\|f_{\res}(X)\|_2^2 \leq C_{\text{aux}}$ for some constant $C_{\text{aux}} > 0$.
\end{Assumption}

This assumption posits that the target subject's true function can be approximated by a linear combination of source models plus a bounded residual. In other words, the target subject shares similarities with the source subjects, plus idiosyncratic structure captured by $f_{\res}$. If the source models are highly informative, then $\E\|f_{\res}(X)\|_2^2$ is small.

\begin{Assumption}[Restricted eigenvalue] \label{assump:RE}
Let $S^* = \{k : \gamma_k^* \neq 0\}$, and $s^* = |S^*|$. Let $F_i = (f_{\widetilde{\Theta}_1}(X_i), \ldots, f_{\widetilde{\Theta}_K}(X_i))^T \in \mathbb{R}^{K \times m}$, and $\widehat{\Sigma} = n^{-1} \sum_{i=1}^n F_i F_i^T$. Suppose $v^T \widehat{\Sigma} v / \| v_{S^*}\|_2^2 \geq \kappa$, for some constant $\kappa > 0$, and all $v \in \Rbb^K$ with $\|v_{S^{*c}}\|_1 \leq 3\|v_{S^*}\|_1$.
\end{Assumption}

This assumption requires that the source model predictions are not too collinear. It is a standard assumption in high-dimensional regressions that ensures consistent estimation and support recovery of Lasso \citep{bickel2009simultaneous}, and is known to hold with a high probability under sub-Gaussian random designs \citep{raskutti2010restricted}.

Next, we establish a finite-sample generalization bound that explicitly quantifies how auxiliary source data enhances learning through the transfer learning mechanism. Let $\Theta'^*_{\res} = (\theta'^*_{L,\res}, \ldots, \theta'^*_{0,\res}) = \arg\min_{\Theta\in\mathcal{M}_q}\E\left\|\left\{ f^*(X) - \sum_{k=1}^K \gamma_k^* f_{\widetilde{\Theta}^{(k)}}(X) \right\} - f_{\Theta}(X) \right\|_2^2$ denote the population risk minimizer corresponding to Step 2 of MTL.

\begin{Theorem}[Generalization bound for MTL] \label{thm:mtl}
Suppose Assumptions \ref{assump:quadratic-growth}, \ref{assump:auxiliary}, and \ref{assump:RE} hold. Suppose $q \in [1,2]$, $\lambda'_{\wts} \asymp \sigma \sqrt{\log K / n}$, $\lambda'_{res} \asymp v_\infty\sqrt{L\{\log(nmp_{\tot})\}^3 / n}$. Then, with probability approaching 1 as $n \to \infty$, we have
\begin{align*} \label{eq:mtl-bound}
\E\left\|\widehat{f}_{\MTL}(X) - f^*(X)\right\|_2^2 \lesssim & \underbrace{\inf_{\Theta\in\mathcal{M}_q} \E\left\|f_{\Theta}(X) - \left\{ f^*(X) - \sum_{k=1}^K \gamma_k^* f_{\widetilde{\Theta}^{(k)}}(X)\right\} \right\|_2^2}_{\text{$\mathcal{E}_{\MTL,1}$: Residual space approximation error}} \notag\\
&\quad + \underbrace{\frac{\sigma^2 s^* \log K}{\kappa^2 n}}_{\text{$\mathcal{E}_{\MTL,2}$: Sparse weights learning statistical error}} \notag\\ 
&\quad + \underbrace{v_\infty \sqrt{\frac{L\{ \log(nmp_{\tot}) \}^3}{n}} \|\theta'^{*}_{L,\res}\|_q^q}_{\text{$\mathcal{E}_{\MTL,3}$: Residual learning statistical error}}  
+ \underbrace{v_\infty^2 \frac{L\{\log(nmp_{\tot})\}^3}{n}}_{\text{$\mathcal{E}_{\MTL,4}$: Higher-order term}}.
\end{align*}
\end{Theorem}

Theorem \ref{thm:mtl} shows that MTL achieves a finite-sample bound where the total error decomposes into four terms: three estimation errors reflecting the quality of knowledge transfer from source tasks, the accuracy of identifying relevant source models, the complexity of learning the residual structure, plus a higher-order term. Specifically, the first term, $\mathcal{E}_{\MTL,1}$ measures how well the neural network class $\mathcal{M}_q$ approximates the residual space after removing the contribution of source models. When source models capture substantial information in the target function, this error term decreases. Unlike ISL, for MTL, the quality of source models and the optimal combination weights jointly  determine the residual space approximation error through their ability to approximate the target function. The second term, $\mathcal{E}_{MTL,2}$, quantifies the error incurred in estimating the weights $\gamma$ using Lasso. It scales with the number of truly relevant source models $s^*$, but only the logarithm of the total number of source models $K$, indicating that MTL can leverage a large number of source models. This error also depends on $\kappa^2$ that reflects the difficulty of identifying the true support using Lasso, and $\sigma^2$ that reflects the noise level in the target task. The third term, $\mathcal{E}_{MTL,3}$, is the statistical error in learning the residual function, where the intrinsic complexity is governed by the norm $\|\theta_L'^{*}\|_q^q$. Again, when source models are informative, learning the residual function leads to a smaller error than learning the target function directly. The fourth term, $\mathcal{E}_{MTL,4}$, is a higher-order term that vanishes at the rate $1/n$. We give the proof of Theorem \ref{thm:mtl} and discuss its main challenge in Appendix Section S2.4. 

Finally, we establish the performance improvement and safeguard property of MTL. 

\begin{Theorem}[Performance guarantee of MTL] \label{cor:mtl-safety} 
Suppose the conditions of Theorem \ref{thm:mtl} hold. 
\begin{enumerate}[(a)]
\item (Safety). When $\|\theta'^{*}_{L,\res}\|_q^q \lesssim \|\theta_L^{*}\|_q^q$, MTL is never worse than the baseline method, in that
\begin{equation*}
\E \|\widehat{f}_{\MTL}(X) - f^*(X)\|_2^2 \lesssim \mathbb{E}\|f_{\widehat{\Theta}_{\base}}(X) - f^*(X)\|_2^2 + \frac{\sigma^2 s^* \log K}{\kappa^2 n}.
\end{equation*}

\item (Enhancement). When (i) $s^* \log K \ll n$, and (ii) the source models are informative, in that $\|\theta'^{*}_{L,\res}\|_q^q < \|\theta_L^{*}\|_q^q - c \sqrt{s^* \log K}$, for some constant $c > 0$, then MTL achieves a strict improvement over the baseline method, in that, for some constant $c' > 0$, 
\begin{align*}
\E \|\widehat{f}_{\MTL}(X) - f^*(X)\|_2^2 \lesssim & \; \mathbb{E}\|f_{\widehat{\Theta}_{\base}}(X) - f^*(X)\|_2^2 \\
& \quad - c' \, v_\infty \sqrt{\frac{L\{ \log(nmp_{\tot}) \}^3}{n}} \left( \|\theta_L^{*}\|_q^q - \|\theta'^{*}_{L,\res}\|_q^q \right). 
\end{align*}
\end{enumerate}
\end{Theorem}

Theorem \ref{cor:mtl-safety} quantifies the performance enhancement that MTL achieves over the baseline method, and characterizes the precise conditions under which this enhancement occurs. 

The safety guarantee shows that MTL incurs only a controlled excess risk bounded by the Lasso estimation error $\sigma^2 s^* \log K / (\kappa^2 n)$ when the residual space learning maintains a comparable intrinsic complexity to the original learning, in that $\|\theta'^{*}_{L,\res}\|_q^q \lesssim \|\theta_L^{*}\|_q^q$. When $s^*$ is small, the Lasso error vanishes at the rate $(\log K) / n$, implying that MTL can safely search over the exponentially large number of source models with only a logarithmic cost. This contrasts sharply with naive ensemble methods that would suffer linear dependence on $K$ in the generalization bounds \citep{tsybakov2003optimal}.

The enhancement guarantee quantifies the benefit of MTL, which is similar as that for ISL. Such an enhancement materializes when two conditions hold jointly. The first condition requires that the effective dimension of the transfer learning problem, measured by $s^* \log K$, remains small relative to the sample size $n$. This ensures that the Lasso estimation error does not overwhelm the benefit of complexity reduction in the residual space. The second condition requires that source models provide sufficiently informative information, in that the intrinsic complexity of residual learning and that of direct learning, as measured by $\|\theta'^{*}_{L,\res}\|_q^q$ and $\|\theta_L^{*}\|_q^q$, respectively, differs by a certain amount, so to ensure that the improvement from reduced complexity dominates the cost of estimating the combination weights. When both conditions hold, the gain scales with the square root of the sample size times the path norm reduction, revealing that the benefit of transfer learning compounds with the availability of labeled data in the target task.

\section{Numerical Studies}
\label{sec:numericals}

\subsection{Benchmark data and experiment setup}
\label{sec:data}

We carry out numerical experiments using the benchmark, the Natural Scenes Dataset (NSD), a large-scale 7T human fMRI data designed to bridge cognitive neuroscience and artificial intelligence \citep{allen2022nsd}. It contains fMRI responses from eight healthy adult participants. Four of them, S1, S2, S5, S7, completed the full protocol of 40 scan sessions over the course of one year, while the other four only completed part of scan sessions. Each scanning session consisted of 12 runs, with each run containing 62 to 63 natural image stimulus trials. In total, each participant viewed approximately 10,000 distinct natural images, with each image repeated three times to ensure reliable BOLD response estimation. All visual stimuli were drawn from the Microsoft Common Objects in Context (COCO) dataset \citep{lin2014coco}, which provides a rich and diverse set of natural scenes. Following the usual protocol in recent fMRI-to-image reconstruction literature  \citep{scotti2023mindeye,OzcelikVanRullen2023,huo2024neuropictor}, we focus our experiments on the four participants who completed all scan sessions. These subjects share a common testing set of 982 fMRI-image pairs, and a different training set out of 8,859 fMRI-image pairs. The dataset is publicly available at \url{https://registry.opendata.aws/nsd/}. 

We follow the general encoder-alignment-decoder structure as described in Figure \ref{fig:schematic}. We adopt the same type of encoder and decoder as those used in latent embedding alignment (LEA) of \citet{qian2024lea}, while we keep the pretrained encoder and decoder \emph{frozen} throughout our numerical experiments \emph{without} any fine-tuning. More specifically, for the fMRI encoder, we adopt the architecture from NeuroPictor \citep{huo2024neuropictor}, which is based on a masked autoencoder similar to that used in LEA. Raw fMRI data are first projected onto 2D cortical flatmaps using the fMRI-PTE preprocessing pipeline \citep{chen2023fmripte}, producing a $256 \times 256$ single-channel activation map for each image-fMRI trial. The encoder partitions the flatmaps into patches, processes them with a vision transformer, and uses a dedicated guide token to aggregate global neural activity into a single 2048-dimensional latent representation. This encoder was pretrained on large-scale UK Biobank fMRI data from over 100,000 participants. For the natural image encoder, we use the ViT-H/14 model from the OpenCLIP library \citep{cherti2023reproducible}. It employs a vision transformer with a $14 \times 14$ patch size and produces a 1024-dimensional embedding for each image. The encoder was pretrained on the LAION-5B dataset containing 5.85 billion image-text pairs. For the natural image decoder, we adopt a CLIP-conditioned MaskGIT architecture built on a VQGAN tokenizer, which defines a discrete visual codebook. Given the predicted embedding, the frozen MaskGIT model autoregressively samples quantized visual tokens, which are then decoded into RGB images by the VQGAN decoder.

\subsection{Methods for comparison and evaluation metrics}
\label{sec:baseline}

Since we focus on the alignment step, we compare our approach with several alternative alignment methods, including ridge regression as used in LEA \citep{qian2024lea}, partial least squares (PLS), and contrastive learning (CL). For CL, we implement a standard CLIP-space contrastive framework in which a two-hidden-layer MLP, with 2048 nodes per layer, maps fMRI latent features to the CLIP image embedding space. The model is trained using a symmetric InfoNCE-style contrastive loss combined with a feature regression term, a common strategy for stabilizing multimodal representation alignment \citep{oord2018cpc, Radford2021CLIP}. We exclude canonical correlation analysis due to its instability and tendency to overfit in high-dimensional settings, and optimal transport because it aligns distributions rather than learning the explicit predictive mapping required for fMRI-to-image reconstruction. In addition, we compare with MindEye, a state-of-the-art fMRI-to-image reconstruction method that maps fMRI signals into the CLIP space, aligns the predicted embeddings with the CLIP image distribution via a diffusion prior, and reconstructs images using a CLIP-conditioned diffusion decoder \citep{scotti2023mindeye}.

We evaluate reconstruction quality using three numerical metrics, along with visual examination. The first metric is the CLIP distance, which quantifies semantic similarity between reconstructed and true images in the CLIP feature space. Specifically, for each image pair, features are extracted with a pretrained CLIP ViT-H/14 model, $\ell_2$-normalized, and compared using cosine similarity:
\begin{equation*}
\text{CLIP Distance} = \frac{1}{n_{\mathrm{test}}} \sum_{i=1}^{n_{\mathrm{test}}} \frac{f_{\text{CLIP}}(\hat{r}_i) \cdot f_{\text{CLIP}}(r_i)}{||f_{\text{CLIP}}(\hat{r}_i)|| \cdot ||f_{\text{CLIP}}(r_i)||}
\end{equation*}
where $f_{\text{CLIP}}(\cdot)$ denotes the CLIP feature extraction function, $\hat{r}_i$ and $r_i$ are the reconstructed and true images, respectively. The second metric is the CLIP correlation, which evaluates whether a reconstructed image is most strongly associated with its corresponding true image. Specifically, for each reconstruction, we compute CLIP feature correlations with all test images and report the proportion of cases in which the correlation with the correct image exceeds that with all others, i.e.,
\begin{equation*}
\text{CLIP Correlation} = \frac{1}{n_{\mathrm{test}}} \sum_{i=1}^{n_{\mathrm{test}}} \frac{1}{n_{\mathrm{test}}-1} \sum_{j \neq i} \mathbb{I}\big[ \rho(f_{\text{CLIP}}(\hat{r}_i), f_{\text{CLIP}}(r_i)) > \rho(f_{\text{CLIP}}(\hat{r}_i), f_{\text{CLIP}}(r_j)) \big],
\end{equation*}
where $\rho(\cdot, \cdot)$ denotes the Pearson correlation, and $\mathbb{I}[\cdot]$ is the indicator function. The third metric is top-$k$ classification accuracy, which evaluates whether the reconstructed image preserves the semantic category of the original image. Because NSD lacks explicit class labels, we assign labels to true images using a fixed pretrained ViT-H/14 classifier and augment them with randomly sampled distractor classes. Reconstructed images are labeled using the same classifier, and accuracy is defined as the fraction of cases in which the predicted label appears in the true top-$k$ set. All three metrics are standard in natural image reconstruction studies \citep{Gaziv2022SelfSupervised, chen2023seeingbeyond, qian2024lea}, and higher values indicate better reconstruction quality. We also report the number of trainable parameters for each method as a measure of computational complexity.

\subsection{Semi-supervised learning}
\label{sec:num-semi}

For semi-supervised learning, we use all 8,859 fMRI-image pairs in the training set as the labeled data, and randomly sample 10k or 50k images from the COCO dataset, excluding the ones overlapping with NSD to prevent data leakage, as the additional unpaired data. We employ an MLP model with three hidden layers of 512, 512 and 256 hidden nodes for the inverse mapping step, and an MLP with two hidden layers of 512 and 256 hidden nodes for the augmented learning and residual learning steps. We carry out the analysis for one subject at a time, then average the results across the four subjects with complete scan sessions. We evaluate the  reconstruction using 982 fMRI-image pairs in the testing set. 

\begin{table}[t!]
\centering
\begin{tabular}{lccccc}
\hline
Method & CLIP Dist & CLIP Corr & Top1 Acc & Top3 Acc \\ \hline
LEA & 0.448$\pm$0.006 & 0.886$\pm$0.008 & 0.364$\pm$0.016 & 0.570$\pm$0.020 \\
PLS & 0.453$\pm$0.005 & 0.892$\pm$0.008 & 0.379$\pm$0.015 & 0.587$\pm$0.018 \\
CL & 0.471$\pm$0.008 & 0.897$\pm$0.010 & 0.427$\pm$0.021 & 0.618$\pm$0.024 \\
MindEye & 0.561$\pm$0.012 & 0.962$\pm$0.004 & 0.614$\pm$0.016 & 0.767$\pm$0.016 \\
ISL (0) & 0.468$\pm$0.008 & 0.885$\pm$0.011 & 0.416$\pm$0.020 & 0.599$\pm$0.023 \\
ISL (10k) & 0.485$\pm$0.007 & 0.898$\pm$0.008 & 0.454$\pm$0.019 & 0.639$\pm$0.021 \\
ISL (50k) & 0.486$\pm$0.008 & 0.900$\pm$0.009 & 0.459$\pm$0.023 & 0.642$\pm$0.025 \\ \hline
\end{tabular}
\caption{Average evaluation metrics across four subjects.}
\label{tab:semi_avg}
\end{table}

\begin{figure}[t!]
\centering
\includegraphics[width=0.95\textwidth]{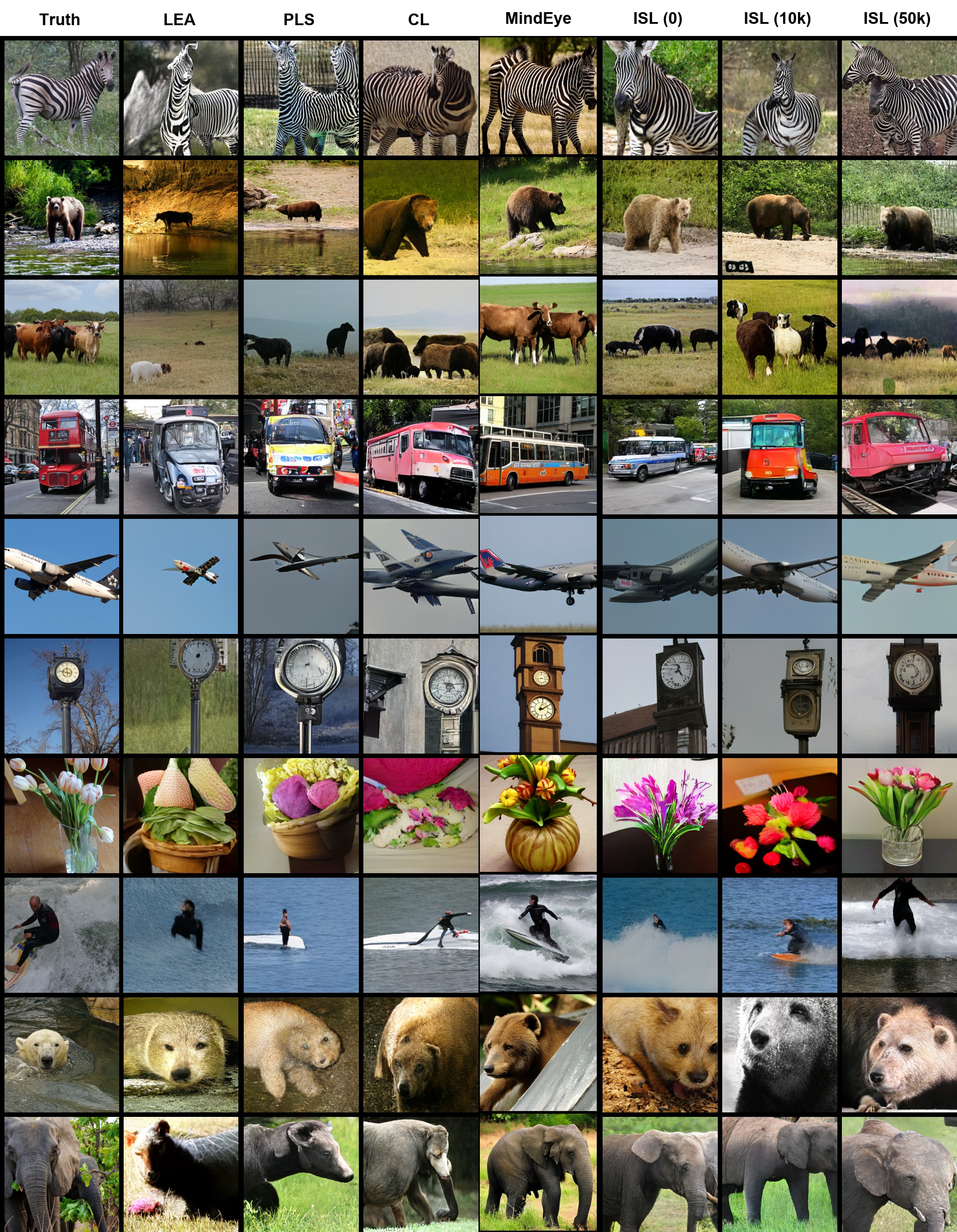}
\caption{Semi-supervised learning reconstruction of NSD Subject S1.}
\label{fig:semi_sbj1}
\end{figure}

Table \ref{tab:semi_avg} reports the average evaluation metrics across four subjects, and Figure \ref{fig:semi_sbj1} reports the reconstructed images for subject S1 with various alignment methods. In the interest of space, we report the visual reconstruction results for the other three subjects in the Appendix Section S3.1. From the table, we see that incorporating additional unpaired images improves alignment performance. In addition, our method outperforms alternative alignment approaches including LEA, PLS, and CL. Although MindEye achieves higher quantitative accuracy, this difference is expected, as MindEye jointly fine-tunes the entire pipeline of encoding, alignment, and decoding, whereas our method focuses exclusively on alignment, with both the encoder and decoder fully frozen. Moreover, Figure \ref{fig:semi_sbj1} shows that, despite the lower numerical metrics, our method, when trained with 50k unpaired images, produces reconstructions that are visually comparable to those of MindEye. Finally, in terms of trainable parameters, LEA, PLS, CL, MindEye, and ISL involve 2.1M, 4.2M, 10.5M, 44M, and 4.3M parameters, respectively. Our method therefore uses roughly one-tenth the number of parameters of MindEye. Taken together, it highlights a clear trade-off: modestly reduced accuracy in exchange for a substantially lighter, more computationally efficient, and theoretically grounded brain encoding  decoding solution.

\subsection{Transfer learning}
\label{sec:num-transfer}

For transfer learning, we randomly sample a subset of fMRI-image pairs of one subject as the target domain, and use all fMRI-image pairs of the other three subjects as the source domain. We employ an MLP model with two hidden layers of 512 and 256 hidden nodes for both the sparse source weight learning and the residual learning steps. 

\begin{table}[b!]
\centering
\begin{tabular}{lcccc}
\hline
\# Images \& Method & CLIP Dist & CLIP Corr & Top-1 Acc & Top-3 Acc \\ \hline
3k (transfer) & 0.456$\pm$0.006 & 0.869$\pm$0.009 & 0.388$\pm$0.018 & 0.569$\pm$0.021 \\
5k (no transfer) & 0.456$\pm$0.006 & 0.873$\pm$0.009 & 0.383$\pm$0.016 & 0.566$\pm$0.018 \\
6k (no transfer) & 0.459$\pm$0.007 & 0.876$\pm$0.010 & 0.389$\pm$0.018 & 0.570$\pm$0.021 \\
\hline
4k (transfer) & 0.468$\pm$0.007 & 0.881$\pm$0.009 & 0.412$\pm$0.020 & 0.596$\pm$0.024 \\
5k (transfer) & 0.473$\pm$0.006 & 0.884$\pm$0.008 & 0.419$\pm$0.015 & 0.603$\pm$0.018 \\
8.8k (no transfer) & 0.468$\pm$0.008 & 0.883$\pm$0.011 & 0.415$\pm$0.020 & 0.599$\pm$0.022 \\
8.8k (transfer) & 0.481$\pm$0.006 & 0.895$\pm$0.008 & 0.443$\pm$0.016 & 0.630$\pm$0.019 \\ \hline
\end{tabular}
\caption{Transfer learning and baseline performance averaged across 4 subjects.}
\label{tab:transfer_avg}
\end{table}

\begin{figure}[t!]
\centering
\includegraphics[width=0.8\textwidth]{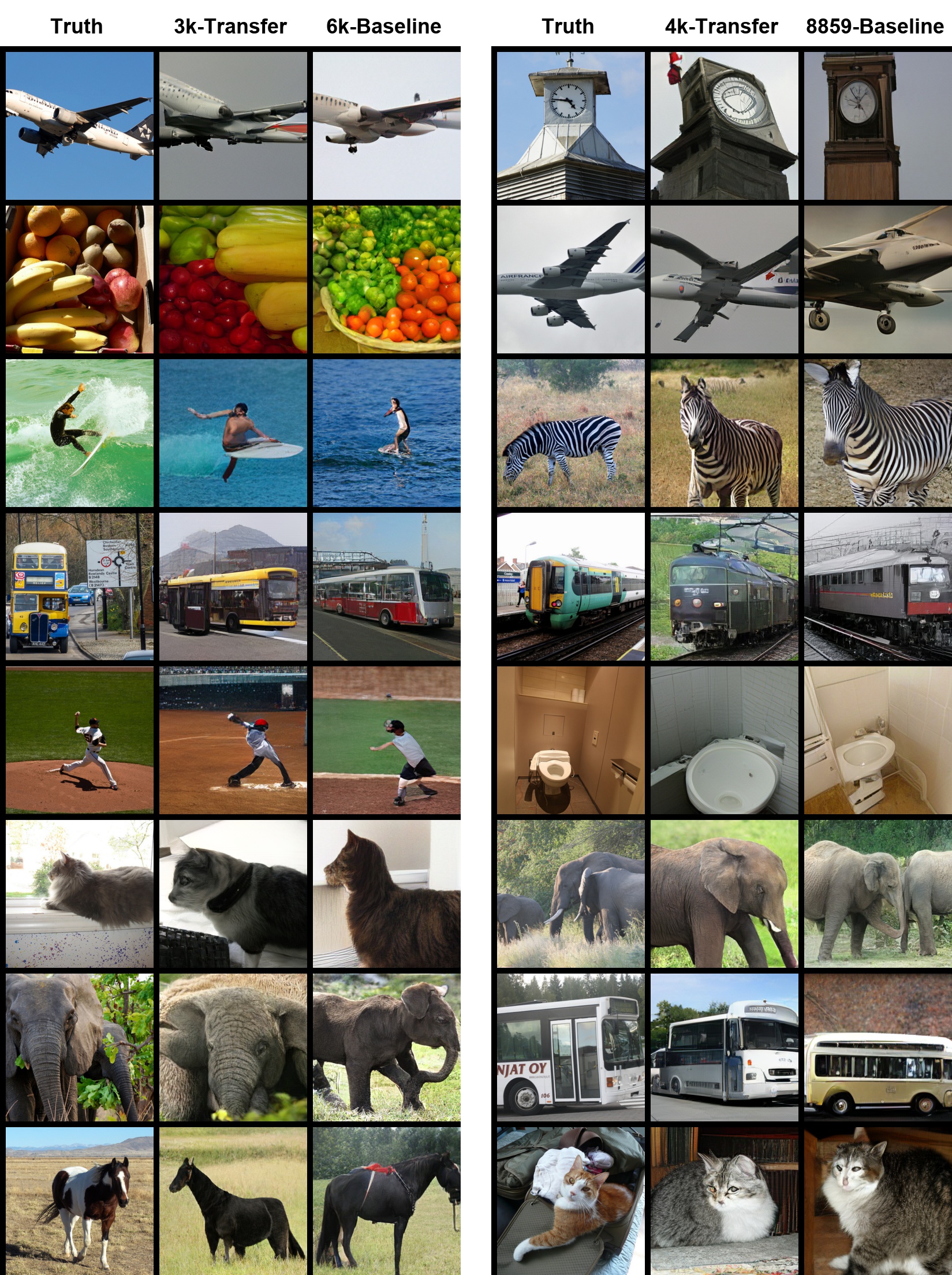}
\caption{Transfer learning reconstruction of NSD Subject S1.}
\label{fig:transfer_sbj1}
\end{figure}

Table \ref{tab:transfer_avg} reports the average evaluation metrics across four subjects, and Figure \ref{fig:transfer_sbj1} reports the reconstructed images for subject S1 with or without transfer learning. Again, we report  the results for the other three subjects in the Appendix Section S3.2. From the table and figures, we see that, comparable reconstruction accuracy can be achieved using roughly half as many fMRI-image training pairs. For instance, using 3k pairs under transfer learning yields performance similar to using 5k to 6k pairs without transfer learning, while using 4k to 5k pairs with transfer learning are comparable to using all 8.8k pairs without transfer learning. Moreover, applying transfer learning with the full set of 8.8k training pairs further improves performance. Together, these findings demonstrate that our transfer learning approach effectively leverages information from other subjects to substantially reduce the training data requirements for a new subject, offering clear practical benefits.

\section{Discussions}
\label{sec:discussions}

In this article, we have proposed two statistical approaches, inverse semi-supervised learning and meta transfer learning, for latent embedding alignment in brain encoding and decoding. Both methods achieve performance improvement through principled complexity reduction, rather than architectural expansion or joint retraining, but differ in how auxiliary information is exploited. In both methods, performance gains arise from reducing the effective complexity of the learning problem through residual-space decomposition, as reflected by a smaller $\ell_q$ path norm governing statistical error. However, ISL achieves this reduction by increasing the effective sample size using unlabeled outputs, relying on inverse mapping estimation and residual correction to control bias introduced by pseudo-predictors. In contrast, MTL reduces complexity by transferring information from related source tasks, using sparse aggregation of pretrained models to capture shared structure across subjects, followed by residual learning to account for subject-specific effects. As a result, ISL is most effective when unlabeled stimulus embeddings are abundant and the inverse mapping is reasonably accurate, whereas MTL is most effective when reliable source models from similar tasks or subjects are available but labeled data in the target task are limited. 

It is useful to extend the proposed framework to incorporate additional modalities, particularly text information associated with visual stimuli. Modern vision-language models encode rich semantic structure through joint image-text embeddings, and textual descriptions may provide complementary information that further regularizes the alignment between neural and stimulus representations. Developing statistically principled methods that exploit such multimodal information, while preserving the safety guarantees, interpretability, and computational efficiency of the current framework, remains an open and promising direction for future research.

\bibliographystyle{apa}
\bibliography{ref-align.bib}

\end{document}